\documentclass{article}
\usepackage{spconf,amsmath,graphicx, amssymb,multirow,booktabs}
\usepackage{caption} 
\title{Waveform Boundary Detection for Partially Spoofed Audio}
\name{Zexin Cai$^{1, \ast}$\thanks{$^\ast$Authors contributed equally}, Weiqing Wang$^{1, \ast}$, Ming Li$^{1, 2, \dagger}$\thanks{$^\dagger$Corresponding author: Ming Li} }
\address{$^{1}$Department of Electrical and Computer Engineering, Duke University, Durham, NC 27708, USA\\
        $^{2}$Data Science Research Center, Duke Kunshan University, Kunshan 215316, PR China \\
        {\small \tt ming.li369@duke.edu} }
\begin{document}
\ninept
\maketitle
\begin{abstract}
The present paper proposes a waveform boundary detection system for audio spoofing attacks containing partially manipulated segments. Partially spoofed/fake audio, where part of the utterance is replaced, either with synthetic or natural audio clips, has recently been reported as one scenario of audio deepfakes. As deepfakes can be a threat to social security, the detection of such spoofing audio is essential. Accordingly, we propose to address the problem with a deep learning-based frame-level detection system that can detect partially spoofed audio and locate the manipulated pieces. Our proposed method is trained and evaluated on data provided by the ADD2022 Challenge. We evaluate our detection model concerning various acoustic features and network configurations. As a result, our detection system achieves an equal error rate (EER) of 6.58\% on the ADD2022 challenge test set, which is the best performance in partially spoofed audio detection systems that can locate manipulated clips. 
\end{abstract}
\section{Introduction}
\label{sec:intro}

Deepfake audios refer to those altered or generated utterances that aim to fool human observers and machines. Deep learning frameworks have significantly improved the speech quality of text-to-speech (TTS) and voice conversion (VC), making the synthesized utterances too real to be distinguished from natural ones~\cite{oord2016wavenet, shen2018natural, kong2020hifi}. In addition, the robustness accomplished by zero-shot synthesis and any-to-any VC approaches makes the voice cloning performance much more powerful~\cite{chou19_interspeech, Cai2020, lin2021fragmentvc, haozhe2021sigvc}. Those approaches achieve the conversion from any voice to another arbitrary voice. Under this circumstance, high-fidelity synthesis/conversion systems inevitably allow criminals to commit fraud by impersonating others. Thus, concerning social security, detecting spoofing utterances is critical to reduce the threat yielded by disinformation embedded in speech utterances. On the other hand, spoofing detection techniques can address the vulnerabilities of automatic speaker verification (ASV) systems against spoofing attacks~\cite{wu2015spoofing, nandwana20_odyssey}.

Regarding threats revealed by audio spoofing attacks, the ASVspoof challenge has been held biennially since 2015 for countermeasures research against various spoofing attacks, including synthetic speech, voice conversion, replay, and impersonation~\cite{kamble2020advances}. Motivated by the challenge, researchers have proposed different advanced deep-learning structures for audio spoofing attacks. For example, Light-CNN (LCNN)~\cite{wu2018light} and RawNet~\cite{tak2021end} have been adopted as the backbone network in the anti-spoofing task and achieved satisfactory performances~\cite{wang21_asvspoof}. Even though many spoofing scenarios are included in the ASVspoof challenge, the spoofing scenario caused by partially spoofed/fake clips is not included. Spoofing utterances under this scenario are generated by altering the original bona fide utterances with natural or synthesized audio~\cite{9746939}. One example is that attackers can change a few words with synthesized clips and ultimately reverse the message carried by the utterance.  Concerning such attacks, two datasets, Half-Truth and PartialSpoof, along with their benchmark systems, are developed to advance the research in partially fake spoofing detection~\cite{yi21_interspeech, zhang21ca_interspeech, zhang21_asvspoof}. 

The recent Audio Deep Synthesis Detection (ADD) challenge has included the task of detecting partially spoofed utterances as one of the challenge tracks~\cite{9746939}. Lv et al. participated in the challenge and achieved the best performance with fake audio detection systems finetuned from unsupervised pretraining models~\cite{9747605}. However, defining the task as a binary classification problem under the utterance level, their proposed system has limitations in finding fake clips in a given utterance. Alternatively, Wu et al. incorporated a question-answering strategy in framework design, which allows the detection system to locate fake regions~\cite{9746162}. Typically, the partially fake audios include some artifacts, like the discontinuity between concatenated segments, that allow us to locate the fake utterance clips. Nonetheless, the best performance from~\cite{9746162} on ADD evaluation datasets needs further improvement compared to the detection system from~\cite{9747605}.

In this paper, we propose a frame-level boundary detection system to detect partially fake audio. While previous approaches for this scenario mostly focus on utterance-level detection, our proposed system utilize the discontinuity between segments and is designed to find the concatenation boundary from the acoustic information under the frame level. Therefore, the system is able to locate the fake utterance pieces from partially fake audio by predicting the probabilities of being a concatenation boundary for each speech frame. In addition, the maximum probabilities of all frames can be considered as the utterance-level score for partially fake audio detection. Specifically, our proposed system employs the Wav2Vec~\cite{wav2vec2} as the feature extractor and the ResNet-1D as the main framework, followed by a Transformer encoder-based frame-level backend classifier for boundary detection. We use data simulated from datasets provided by the ADD challenge for training. Our proposed method is evaluated and analyzed with multiple test sets, including out-of-domain data. Results show that the proposed system outperforms other partially fake audio detection systems with boundary detection. It achieves an EER of 3.14\% on the ADD adaptation set and an EER of 6.58\% on the ADD test set.

\begin{figure*}[ht]
  \centering
  \includegraphics[scale=0.6]{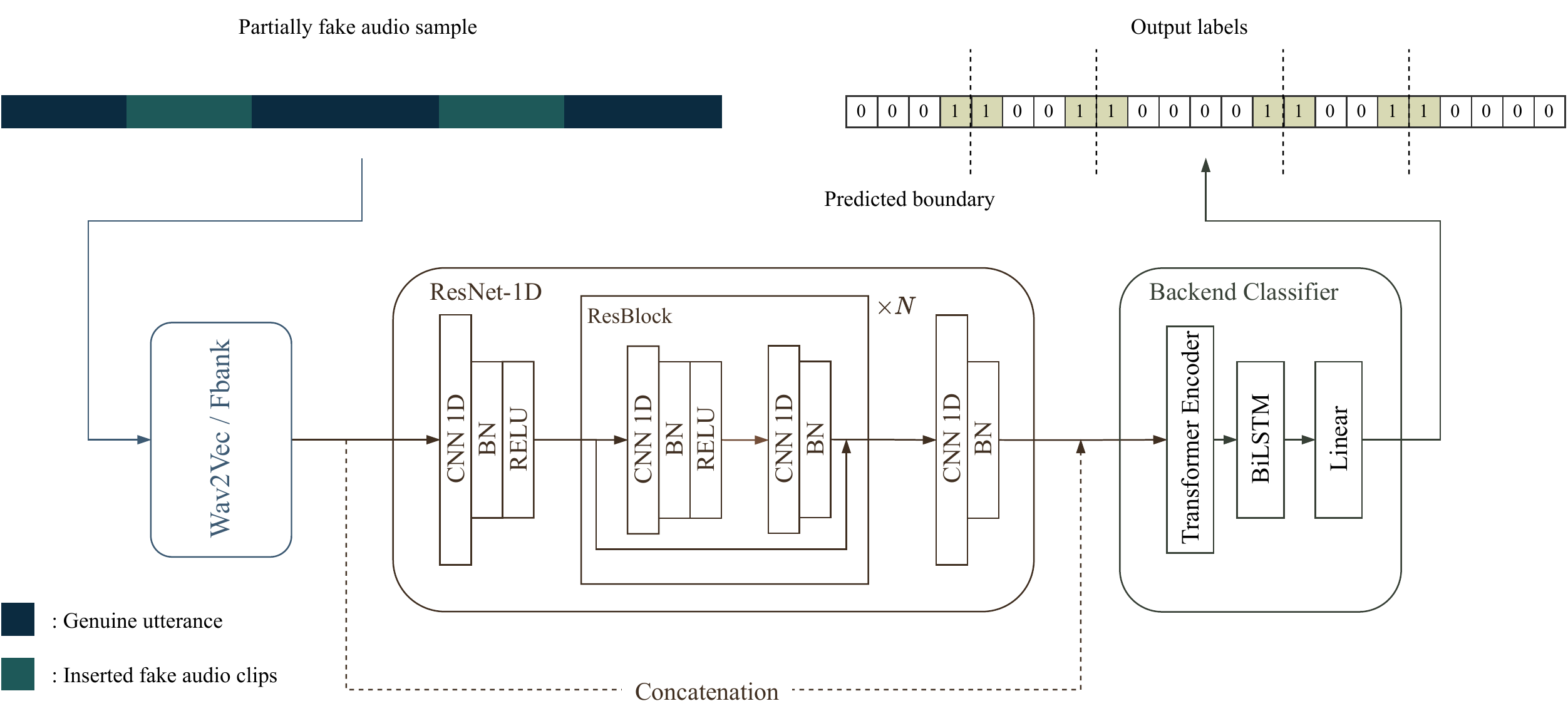}
  \caption{The architecture of our proposed model.}
  \label{fig:architecture}
\end{figure*}

The rest of the paper is organized as follows: In Section~\ref{sec:method}, we present our proposed method for detecting partially spoofed audios. Our experimental setups and inference process are elaborated in Section~\ref{sec:exp}. Then we present and analyse our experimental results in Section~\ref{sec:discuss}. Our work is concluded in Section~\ref{sec:conclu}.

\section{Method}
\label{sec:method}

\subsection{Task Definition}
Considering an audio signal that contains fake audio segments, our purpose is to detect the frames that contain the information of discontinuity. Given the acoustic feature input $\mathbf{X} = (\mathbf{x}_1, \mathbf{x}_2, ..., \mathbf{x}_T) \in \mathbb{R}^{D \times T}$ where $D$ is the feature dimension and $T$ is the number of frames, this task can be considered as a frame-level binary classification task with labeling information $\mathbf{y} = (y_1, y_2, ..., y_{T}) \in \{0, 1\}^{T}$. 

Unlike other methods that only predict utterance-level binary decision~\cite{9747605}, our proposed system can detect if a frame contains such discontinuity. If a frame is a boundary between fake audio and genuine audio, the label is set to 1; otherwise, it is 0. In addition, the labels of frames near the boundary are also set to 1 to increase the robustness. The proposed method can not only find if an audio signal contains fake audio segments but can also identify where the segment is. 

\subsection{Model Architecture}
Figure~\ref{fig:architecture} shows the architecture of our proposed model. We employ a pre-trained Wav2Vec2~\cite{wav2vec2} to extract the features from raw audio. Next, we employ a ResNet-1D to further extract the frame-level embeddings. Then we concatenate the acoustic feature and the frame-level embeddings together as the input of a Transformer-based frame-level classifier, which outputs the probability of being a boundary for each frame.

\subsubsection{Feature extraction}
Wav2Vec is a self-supervised model which learns representations from raw audio data~\cite{wav2vec, wav2vec2}. It has shown impressive performance on many downstream tasks like automatic speech recognition (ASR)~\cite{baevski2021unsupervised}. Therefore, we also employ this unsupervised model as a feature extractor for raw audio samples. More specifically, we employ the Wav2Vec 2.0 (wav2vec2-base-960h\footnote{https://huggingface.co/facebook/wav2vec2-base-960h}). The model wav2vec2-base-960h is trained on the Librispeech dataset~\cite{panayotov2015librispeech}, which contains 960 hours of speech utterances. The output shape and the number of parameters of this model are shown in Table~\ref{tab:wav2vec}. The frame rate of the output feature is 20ms. On the other hand, we also use the Mel filterbank (Fbank) feature for comparison. We extract the 80-dimensional Fbank as well as 1st- and 2nd-order deltas with frame length of 25ms and frame shift of 10ms. Thus the total dimension of the final Fbank feature is 240 and the frame rate is 10ms when using Fbank feature.

\begin{table}[htb]
  \caption{Acoustic feature size}
  \label{tab:wav2vec}
  \centering
  \begin{tabular}[c]{cccc}
    \toprule
    \textbf{Feature}  & Dimension & Frame rate & \#Parameters\\
    \midrule
    Wav2Vec & 768 & 20ms & 95m \\
    Fbank & 240 & 10ms & - \\
    \bottomrule
  \end{tabular}
\end{table}

\subsubsection{Frame-level embedding extraction}

We employ a ResNet-1D network after the feature extractor to obtain frame-level embeddings. The architecture is shown in Figure~\ref{fig:architecture}. The ResNet-1D contains two CNN-1D layers sandwiching N residual blocks, each of which contains two CNN-1D layers and a residual connection from input to output. 

Give the input feature $\mathbf{X} \in \mathbb{R}^{D \times T}$, the ResNet-1D extracts a frame-level embedding sequence $\mathbf{S}\in \mathbb{R}^{D^{\prime} \times T}$, where $D$ is the feature size, $D^{\prime}$ is the frame-level embedding size, and $T$ is the number of frames. 

\subsubsection{Frame-level classifier}
To capture the long-range global context within frames, we employ several Transformer encoders. Later, we use a BiLSTM to further model the sequential embedding from Transformer encoders, and finally, a fully connected layer predicts the boundary probability for each frame.

\section{Experiment}
\label{sec:exp}

\subsection{Data Preparation}
\label{sec:dataprep}
Our training data is constructed from several datasets provided by the ADD challenge~\cite{9746939}. Table~\ref{tab:dataset} presents the statistics of datasets used in our experiments. As shown in Table~\ref{tab:dataset}, the ADD-train dataset contains 3012 genuine utterances and 24072 fake utterances, while the ADD-dev dataset contains 2307 genuine utterances and 26017 fake utterances. All fake utterances from ADD-train and ADD-dev are synthesized by mainstream speech synthesis and voice conversion systems. The ADD-adaptation dataset has 1052 partially fake utterances, and the ADD-test consists of 100625 utterances without labeling information. 
\begin{table}[h]
  \captionsetup{justification=centering}
  \caption{The statistics of datasets (\#Utterances)}
  \label{tab:dataset}
  \centering
  \begin{tabular}[c]{cccc}
    \toprule
    \textbf{Name}  & Genuine & Fake & All \\
    \midrule
    ADD-train & 3012 & 24072 & 27084 \\
    ADD-dev & 2307 & 21295 & 23602 \\
    ADD-test & - & - & 100625 \\
    ADD-adaptation & 0 & 1052$^\ast$ & 1052 \\
    Partially-fake & 0 & 35808$^\ast$ & 35808 \\
    \bottomrule
    \multicolumn{4}{l}{$^\ast$ denotes partially fake utterances}
    
  \end{tabular}
\end{table}
The Partially-fake dataset is generated from the ADD-train dataset. We first use a speech recognition model trained with AISHELL-2~\cite{du2018aishell} from Kaldi~\cite{povey2011kaldi} to obtain the transcript and word boundary information of the ADD-train dataset. As we have the word boundary information of each utterance, we insert audio clips into the genuine utterances according to the following strategies:

\begin{enumerate}
  \item For each genuine utterance, randomly replace $n\in \{1,2,3\}$ word segments with word segments from other genuine utterances. 
  \item For each genuine utterance, randomly replace $n\in \{1,2,3\}$ word segments with word segments from fake utterances.
  \item For genuine utterance $i$, randomly repeat $n \in [1, \lfloor N_i^w/3 \rfloor]$ word segments, where $N_i^w$ is the number of word segments of the utterance. 
\end{enumerate}
Every strategy is applied to every utterance several times to construct partially fake utterances. Correspondingly, we generate 35808 partially fake utterances for training. As for model training, we combine the genuine utterances from the ADD-train dataset and the Partially-fake dataset as the final training data. Besides the ADD-test set, we randomly select 1052 utterances from the ADD-dev set and combine them with the ADD-adaptation set as the adaptation dataset for evaluation. 

\begin{table}[h]
    \caption{The network architecture, where $\mathbf{C}$(kernal size, padding, stride) denotes the convolutional layer, $\left[\cdot \right]$ denotes the residual block, $\mathbf{E}$(layers, heads, FFN size) denotes the Transformer Encoder, BiLSTM(layers, hidden units) denotes BiLSTM layer, Linear(input size, output size) denotes the fully-connected layer; $L$ relates to the duration of the speech and $T$ is the number of label frames. }\

    \centering
    \begin{tabular}[c]{@{\ \ }l@{\ \ }c@{\ \ }l@{\ \ }}
        \toprule
        \textbf{Layer} & \textbf{Output Size} & \textbf{Structure} \\
        \midrule
        Input audio & $L \times 1$ & - \\
        \midrule
        Wav2Vec & $T \times 768$ & - \\
        or Fbank & $T \times 240$ & - \\
        \midrule
        CNN 1D & $T \times 512$ & $\mathbf{C}(5, 2, 1) \mathrm{w/o\ bias}$\\
        \midrule
        ResBlock(s) & $T \times 512$ & $\begin{bmatrix}
            \mathbf{C}(1, 0, 1)\ \mathrm{w/o\ bias} \\
            \mathbf{C}(1, 0, 1)\ \mathrm{w/o\ bias}
        \end{bmatrix}\times12$ \\
        \midrule
        CNN 1D & $T \times 128$ & $\mathbf{C}(1, 0, 1)$\\
        \midrule
        Transformer Encoder & $T \times 128$ & $\mathbf{E}(2, 4, 1024)$\\
        \midrule
        BiLSTM & $T \times 128$ & BiLSTM(1, 128)\\
        \midrule
        Linear & $T \times 1$ & Linear(256, 1)\\
        \bottomrule
    \end{tabular}
    \label{table:architecture}
\end{table}

\subsection{Model Configuration}
Table~\ref{table:architecture} shows the architecture of the proposed model. ResNet-1D contains 12 Residual blocks, each CNN-1D in the residual block has no bias with kernel size=1, and the input and output sizes are 512. The first CNN-1D has no bias with kernel size=5. The input and output sizes are 768 and 512, respectively. The kernel size of the final CNN-1D is 1. The input size is 512, the output is the frame-level embedding, and the embedding size is set to 128.

For the frame-level classifier, the Transformer encoder contains four heads and two encoder layers. The size of the feed-forward network (FFN) is 1024. The BiLSTM contains 128 hidden neurons followed by a ReLU activation. Finally, a 256-d fully-connected layer predicts the probability for each frame.

\subsection{Training Process}
During training, we randomly pick an audio sample from the Partially-fake dataset and the genuine data from ADD-train. The probability of selecting a positive sample is set to 0.5 to ensure data balance. Each audio is cut to a fixed length $l$, e.g., 0.64s, 1.28s, 2.56s. MUSAN~\cite{MUSAN} and RIRs~\cite{RIRs} corpus are employed for online data augmentation. For positive samples, we set the labels to zero vectors. For negative samples, the labels of the boundaries between genuine audio and fake audio are set to 1 and others are set to zeros. In addition, we also set the labels near boundaries to 1. In our experiments, we set labels of 4 closest frames to 1 for each corresponding boundary. 

The model is trained with binary cross entropy loss and Adam optimizer for 100 epochs. The mini-batch size is set to 64. The learning rate is set to $10^{-4}$, and Noam scheduler~\cite{vaswani2017attention} with 1600 warm-up steps is employed. During training, we evaluate the equal error rate (EER) on the adaptation set for validation, and the five models with the lowest EER will be averaged for inference and evaluation.

\subsection{Inference Process and Evaluation}
During the inference stage, we first split each audio signal into several pieces with overlap. The length of each piece is the same as that of the training samples. The network can predict probabilities for each frame of each piece. Next, for each audio signal, we merge the results of all pieces by averaging the overlapped regions. The frames with probabilities greater than a threshold are considered the boundaries, and the mean of the largest $n$ probabilities is the utterance level confidence score. In our experiments, $n$ is set to 4 for best performance. 

\section{Results and Discussion}
\label{sec:discuss}

Our first experiment is conducted to investigate our model's performance concerning different waveform lengths  during training. The model trained with Wav2Vec2 feature extractor, but without the concatenation operation of acoustic feature and the embedding vector, is used as the backbone model in this experiment. Results are shown in Table~\ref{tab:len_exp}. The model achieves the best performance when we set the fixed length $l$ to 1.28s during training. With $l$ set to 1.28s, the EER on the adaptation set is 3.71\%, and the performance on the ADD-test set is 6.64\%. Shorter waveform length like 0.64s cause significant performance degradation on evaluation sets, while increasing $l$ to 2.56s results in minor performance degradation.

\begin{table}[h]
  \captionsetup{justification=centering}
  \caption{The performance of Wav2Vec (without feature-embedding concatenation) , reported in EER.}
  \label{tab:len_exp}
  \centering
  \begin{tabular}[c]{cccc}
    \toprule
    \multirow{2}{*}{\textbf{Evaluation Set} } & \multicolumn{3}{c}{\textbf{Wav Length (s)}} \\
    & 0.64 & 1.28 & 2.56 \\
    \midrule
    Adaptation & 4.85\% & \textbf{3.71\%} & 4.09\% \\
    ADD-test & 9.39\% & \textbf{6.64\%} & 6.69\% \\
    \bottomrule
  \end{tabular}
\end{table}

Therefore, we set the fixed waveform length $l$ to 1.28s for model training. Then we conduct experiments to explore our proposed method regarding various acoustic features and the importance of specific network components. As shown in Table~\ref{tab:mdl_exp}, we remove part of the component from our proposed network and report the corresponding performance. The model `w/o concatenation' denotes the model without the concatenation operation of the acoustic feature and the embedding output vector from the ResNet-1D. The model `w/o ResNet-1D' has the frame-level embedding extractor ResNet-1D removed. As shown in Table~\ref{tab:mdl_exp}, the EER on the adaptation set increases significantly as we remove the concatenation operation, while the performance on ADD-test set degrades slightly. However, after removing the ResNet-1D module, the EER on adaptation is 3.23\%, and the EER on ADD-test set is 6.98\%. Note that utterances from the adaptation set are constructed from the same dataset as our training set, while the ADD-test set contains utterances constructed from other data. Therefore, the adaptation set is an in-domain validation set, and the ADD-test contains out-of-domain data. The result indicates that the acoustic feature extracted by Wav2Vec generalizes well for in-domain data. Without the frame-level embeddings from ResNet-1D, the model still achieves an EER of 3.23\% on the adaptation set. In contrast, the feature extracted by the module ResNet-1D works better on the out-of-domain data. We can see that the EER on ADD-test only increase 0.06\%, while the EER on the adaptation set has increased from 3.14\% to 3.71\%. In general, adding the concatenation operation and the ResNet-1D module helps improve the model performance.  

\begin{table}[h]
  \captionsetup{justification=centering}
  \caption{Performances (EER) regarding various feature extractor and architecture designs, w/o means without.}
  \label{tab:mdl_exp}
  \centering
  \begin{tabular}[c]{lcc}
    \toprule
    \textbf{Model} & \textbf{Adaptation} & \textbf{ADD-test} \\
    \midrule
    Wav2Vec & \textbf{3.14\%} & \textbf{6.58\%}  \\
    \text{\ \ \ \ w/o concatenation}  & 3.71\% & 6.64\% \\
    \text{\ \ \ \ w/o ResNet-1D}  & 3.23\% & 6.98\% \\
    \hline
    Fbank & 3.8\% & 15.76\%  \\
    \text{\ \ \ \ w/o ResNet-1D} & 6.46\% & 22.92\% \\
    \hline
    Utterance-level detection~\cite{9747605} & 3.33\%$^\ast$ & \textbf{4.8\%} \\
    Fake Span Discovery~\cite{9746162} & - & 11.1\%\\
    \bottomrule
    \multicolumn{3}{l}{$^\ast$ The number of genuine utterances used for evaluation }\\
    \multicolumn{3}{l}{might be different}
  \end{tabular}
\end{table}

We also try replacing the Wav2Vec feature extractor with the logarithmic Mel-spectrogram (Fbank) feature. For models under this scheme, we exclude the concatenation operation because the Fbank feature is rather a raw acoustic feature from the frequency domain. The performance of this model is compared with the model Wav2Vec without concatenation. Regarding the in-domain adaptation dataset, the performance of Fbank is close to the system using Wav2Vec. However, for the ADD-test set that contains out-of-domain data, the performance of Fbank is much worse than that of Wav2Vec, whereas the Wav2Vec model achieves an EER of 6.64\% and the Fbank-based model achieves 15.76\%. The result demonstrates that the model trained with features extracted by Wav2Vec generalizes better than Fbank in this task. After we remove the ResNet-1D module from the Fbank-based model, the performance degrades significantly on both evaluation sets. This shows that the Fbank is a lower-level acoustic feature compared to those extracted from the Wav2Vec. Thus the ResNet-1D module is much more crucial here to extract the frame-level feature for better performance.

In addition, we compare our results to models in the literature. The utterance-level detection system from~\cite{9747605} also uses the Wav2Vec feature as the acoustic feature. The utterance-level system adopts a pretrained Wav2Vec model with one billion parameters. It achieves an EER of 3.33\% on the adaptation set and 4.8\% on the ADD-test set. Note that the fake utterances in the adaptation set are the same, but the number of genuine utterances for evaluation is not reported in~\cite{9747605}. It is still the best performance on ADD-test set so far. We failed to fit the corresponding Wav2Vec model into our proposed framework as the parameter size is too large for our machine. Nevertheless, our model is designed as a frame-level detection system, which is capable of locating fake clips from partially fake utterances. The fake span discovery model from~\cite{9746162} also can perform frame-level analysis and uncover fake clips. The best single model based on fake span discovery achieves an EER of 11.1\% on the ADD-test set, while our best performance is 6.58\%, which makes our proposed method the best waveform boundary detection system on the ADD-test set. 

\section{Conclusion}
\label{sec:conclu}
We propose a frame-level partially fake audio detection method, which can not only provide an utterance-level binary decision for partially spoofed audio but also predict where the fake clips are inserted or replaced. We take an ablation study on our model components and  evaluate different acoustic features including Wav2Vec and Fbank. Experimental results show that the proposed method has similar performance to the  best utterance-level system on ADD challenge on adaptation set, and it has better performance than other boundary detection systems for partially spoofed audio.


\bibliographystyle{IEEEbib}
\bibliography{refs}

\end{document}